\documentstyle[preprint,aps]{revtex}

\begin{document}
%\draft
%\widetext
\newcommand{\Sp}{\,}

\title{Thermal activation of exclusonic quasiparticles \\
%Duality of quasiparticle gases \\
in the fractional quantum Hall effect} 
\author{J. Shiraishi$^1$ , M. Kohmoto$^1$ and Y.S. Wu$^2$}
\address{$^1$Institute for Solid State Physics, 
University of Tokyo, Roppongi, Minato-ku, Tokyo 106, Japan}
 \address{$^2$Department of Physics, University of Utah,  Salt
Lake City, UT 84112, U.S.A.}

%\date{\today}
\maketitle

%\widetext
\begin{abstract}
\leftskip 54.8pt
\rightskip 54.8pt

Quasiparticles of the fractional quantum Hall systems obey 
fractional (including mutual) exclusion statistics. In this note
we study the effects of exclusion statistics on thermal activation 
of quasiparticle pairs in the approximation of generalized 
ideal gas. The distribution function for thermally activated 
quasiparticles is found explicitly for the statistics matrix 
given by the composite fermion picture and the thermodynamic 
consequences are discussed analytically.
In particular, at low 
temperatures, there is a quasielectron-quasihole duality  in the 
description of the system between two adjacent magic fillings 
$\nu(m,p)\equiv m/(2pm+1)$ and $\nu(m+1,p)$.

\end{abstract}

%%%%%%%%%%%%%%%%%%%%%%%%%%%%%%%%%
%\section{Introduction}
%%%%%%%%%%%%%%%%%%%%%%%%%%%%%%%%%
%\begin{multicols}{2}
%
A fascinating aspect of the fractional quantum Hall 
(FQH) effect is that quasiparticles in the system 
have exotic quantum numbers, very different from
those of the constituent particles,  the electrons.
For example, the gapful FQH liquid with electron 
filling close to $\nu =m/(2pm+1)$  ($p$ and $m$
being integers) has two species  of quasiparticle 
excitations (quasielectrons  labeled by $+$ and 
quasiholes labeled by $-$) \cite{Laughlin}, both 
of  which are fractionally charged  
($e^{*}_{\pm}=\mp 1/(2pm+1)$) anyons 
\cite{Halperin,Arovas} with fractional exchange 
statistics \cite{Wilczek}. Furthermore, 
they satisfy novel quantum statistics 
in the sense of state counting, which is 
neither Bose or Fermi. In this note we study the
effects of quantum statistics on thermal
activation of the FQH quasiparticle pairs and
their thermodynamic consequences.

Previously it was known that a single state 
can not be occupied by more than one fermions;
on the other hand, any number of bosons can 
occupy a single state. Haldane \cite{Haldane} 
first proposed that in some strongly correlated 
many-body systems, the system may be described 
by quasiparticles that obey {\it fractional exclusion 
statistics} other than Bose or Fermi statistics. We 
call such particles or quasiparticles as "{\it exclusons}". 
One of us \cite{Wu} has considered generalized ideal
gas of non-interacting exclusons, and formulated their
statistical distribution and thermodynamics. (Note that 
this does not mean the interactions are neglected. For 
example, there are one-dimensional exactly solvable 
models which can be precisely represented as  a gas 
of non-interacting exclusons; see {\it e.g.} 
\cite{BerWu,Isakov2,HKKW}). In refs. 
\cite{Haldane,HXZhang,Su,CanJohn,SWYang} it is shown that
quasiparticles in the FQH systems with electron filling
$\nu=1/(2p+1)$ obey fractional exclusion statistics,
including mutual exclusion statistics between 
quasielectrons and quasiholes. In ref. \cite{Wu} the 
low-temperature thermodynamics of FQH quasiparticles in 
these systems is discussed in the framework of 
generalized ideal gas. In this note we consider 
the more general FQH systems close to 
electron filling $\nu (m,p)\equiv m/(2pm+1)$. 
One way to study the effects of exotic exclusion 
statistics, especially the effects of  mutual 
statistical exclusion between quasielectrons and 
quasiholes, is to see how  they affect thermal 
activation of quasiparticle pairs. When 
the statistics matrix of quasiparticles is the one
determined by the composite fermion picture 
\cite{Jain}, we have been able to find explicit 
distribution functions for quasiparticles at 
finite temperature and to discuss analytically
the resulting thermodynamics. 
Such an analytic solutions 
have not been available even for one-dimensional 
solvable models.
In particular, 
we find a quasiparticle-quasihole duality in
the quasiparticle gas at low temperatures: A system 
with $\nu$ between two adjacent magic fillings $\nu(m,p)$
and $\nu(m+1,p)$ can be equivalently described either in 
terms of quasielectrons in the FQH fluid with $\nu(m,p)$ 
or in terms of quasiholes in the FQH fluid 
with $\nu(m+1,p)$. 
%The nontrivial feature of our 
%results is that we have included certain mutual
%statistics between quasielectrons and quasiholes. 

In Jain's composite fermion approach,
the fractional quantum Hall (FQH) state 
of electrons in physical magnetic field is 
explained as the integer quantum Hall (IQH)
state of composite fermions in an effective 
magnetic field \cite{Jain}. Imagine an adiabatic
process, by which somehow a certain amount of 
magnetic flux is collected to electrons, so that 
finally $2p$ flux quanta ($p$ a positive even integer) 
are attached to each electron to form a composite 
fermion. These composite fermions are now moving 
in a reduced effective magnetic field  $B_{eff}=
B-4\pi p \rho$, where $\rho$ is the number density of 
electrons (or composite fermions). (In our convention, 
the unit of flux quantum is $2\pi$.) Therefore the effective
filling factor for composite fermion has been increased to
$\nu _{eff}$, given by $\nu ^{-1}_{eff}=
(B-4\pi p \rho)/2\pi \rho =\nu^{-1}-2p$. 
For $\nu _{eff}=m$ (with $m$ an integer), we have
$\nu^{-1}=2p+m^{-1}$ and
$
\nu={m/(2pm+1)}.         
$
Thus, fractional quantum Hall systems with 
$\nu(m,p)\equiv m/(2pm+1)$ are adiabatically
changed into an integer Hall system with 
filling factor $m$, as was emphasized by 
Greiter and Wilczek \cite{Greiter}.

%%%%%%%%%%%%%%
%\section{statistics matrix}
%%%%%%%%%%%%%%%

Let us consider quasiparticles, quasielectrons and quasiholes,
in the  FQH liquid with filling close to 
$\nu_0 =m/(2pm+1)$. Let $N_\phi \equiv eBV/hc$, $N_+$
and $N_{-}$ be the number of total flux quanta, the number 
of quasielectrons and quasiholes, respectively.  Since 
quasielectrons and quasiholes are fractionally
charged  ($e^{*}_{\pm}=\mp 1/|2pm+1|$), 
the total charge balance means
$
N_e=\nu_0 N_\phi+|e^{*}_+| N_+ - |e^{*}_-| N_-,
$
and the filling $\nu$ is given by 
\begin{eqnarray} 
\nu \equiv {N_e \over N_\phi } &=&
\nu_0 + |e^{*}_+| {N_+ \over N_\phi }-|e^{*}_-|
{N_-\over N_\phi }  \\
%&=& \nu_0 +{1\over |2pm+1|}{N_+ \over  N_\phi}-
% {1\over |2pm+1|}{N_- \over  N_\phi} \nonumber \\
&=& {m\over 2pm+1}+{1\over (2pm+1)^2}(n_+-n_-)\;, \nonumber   
\label{flux-constraint}   
\end{eqnarray} 
where $ n_{\pm}$ is the average ``occupation number per state'' 
$
n_\pm = {N_\pm/|e^*_\pm|N_\phi} = {|2pm+1| N_{\pm} / N_\phi} ,        
$
which is related to the density of
quasiparticles $\rho_{\pm}(T)$  by
$
\rho_{\pm}(T) ={|e^*_\pm| N_\phi / V} n_\pm.     
$
In what follows, we 
restrict the range of the filling factor as $ \nu<1/2p$ 
(or $\nu>1/2p$) if $m\geq 0$ (or $m\leq -1$).

Let us count states of quasiparticles in the FQH system
with filling near $\nu_0=m/(2pm+1)$. The effective 
flux for the composite fermions is given by  
$
N_{\phi,eff}= N_{\phi} - 2p N_{e}\, .
$
At $\nu_0$ the ground state of the system corresponds
to $|m|$ filled Landau levels for composite fermions.  
Note that ${\rm sgn}(N_{\phi,eff})={\rm sgn}(m)$.
We will consider the temperature range that is low enough
so that  the quasiparticle excitations of the system are
dominated by composite fermions excited into their 
($|m|+1$)-th Landau level. Thus, quasielectrons 
correspond to composite fermions in the ($|m|+1$)-th 
Landau level, while quasiholes holes of
composite fermions in the filled $|m|$-th Landau level. 
If the composite-fermion excitations, which should be
considered as unit-charged for consistency, obey usual 
Fermi statistics, then the number of available single-particle 
states for them are obviously

\begin{equation}
G_{eff,\pm}^{(b)}=|N_{\phi,eff}| - (N_{\pm}-1) .                
\label{compfermi}
\end{equation}
To derive the exclusion statistics of the 
quasiparticle excitations, we need to express 
$N_{\phi,eff}$ in  (\ref{compfermi}) 
in terms of the external $N_{\phi}$, resulting in

\begin{eqnarray}
G_{eff,+}^{(b)} &=& G_+ - %\frac{1}{|2pm+1|} N_{\phi} -
\alpha_{++} N_{+} 
-\alpha_{+-} N_{-}, \nonumber \\
%\left(1+\frac{2p}{2pm+1} \right) N_{+} 
%+\left(\frac{2p}{2pm+1}\right) N_{-} \nonumber \\
G_{eff,-}^{(b)} &=& G_-  -%\frac{1}{|2pm+1|} N_{\phi} -
\alpha_{-+} N_{+} 
-\alpha_{--} N_{-},
%\left( \frac{2p}{2pm+1}\right) N_{+}
%-\left(1-\frac{2p}{2pm+1}\right) N_{-}\, .
 \label{Jainscheme}
\end{eqnarray}
where $G_\pm=|e^*_\pm | N_{\phi}$ 
is the single-quasiparticle degeneracy in terms
of the external flux
with the proportionality constant $1/|2pm +1|$ identified
as the fractional charge (absolute value) of the  
quasiparticles, and 
the exclusion statistics matrix $\alpha_{\pm\pm}$ for quasiparticles
are:
\begin{eqnarray}
\alpha_{\epsilon \epsilon'}  = \delta_{\epsilon,\epsilon'}
+ \epsilon'{2p\over {2pm+1}}.
\label{jainstat}
\end{eqnarray}
We emphasize that these values of statistics are derived 
under the assumption of eq. (\ref{compfermi}), which 
looks very natural in the composite fermion picture.
(However, modification is not absolutely impossible. We
leave this problem to discussions at the end of the paper.)

%This statistics matrix has several interesting symmetries.
%We do not have no special symmetry for general $p$ and $m$. 
%If we consider the limit $|m|\rightarrow\infty$,
%{\it i.e.} $\nu_0 \rightarrow 1/2p$,  
%the statistics matrix tend to that of free fermion.
%Moreover, we have a particle-hole symmetry which is 
%valid up to the order of $1/m$ 
%\begin{eqnarray}
%&&\alpha_{++}(m)  = \alpha_{--}(-m)+ { O}(1/m^2), \qquad\qquad\;
% \alpha_{+-}(m)= \alpha_{-+}(-m)+ { O}(1/m^2), \nonumber \\
%&&\alpha_{-+}(m)  = \alpha_{+-}(-m)+ { O}(1/m^2), \qquad\qquad 
%\alpha_{--}(m) = \alpha_{++}(-m)+ { O}(1/m^2).   
%\end{eqnarray}
%Finally, 
The diagonal statistics satisfies
$
\alpha_{--}(m+1) ={1/ \alpha_{++}(m)}.
$
This ``symmetry'' becomes important while we study
the  duality in the quasiparticle gas at low temperatures.
%plateau transition at low temperature, 

%%%%%%%%%%%%%%
%\section{thermodynamics of quasi particles}
%%%%%%%%%%%%%

When the temperature is low, the density of thermally
activated quasiparticle pairs is not very high, so that
quasiparticle interactions other than statistical ones
may be ignored  in the first approximation. The system 
is therefore described by a generalized ideal gas 
\cite{Wu} with two species, for which the following 
two conditions are satisfied: 

\noindent
1) The total energy (eigenvalue) 
is always of the form of a simple sum, in which 
the $i$-th term is linear in the particle number 
$N_{i}$:
$ E=\sum_{i} N_{i} \varepsilon_{i}, $

\noindent 
with $\varepsilon_{i}$ identified as the
energy of a quasielectron ($i=+$) or a quasihole ($i=-$).

\noindent
2) The following state-counting for statistical
weight $W$, i.e. the dimension of the Hilbert space
of quasiparticle states, is applicable:

\begin{eqnarray} 
W = 
\left(\begin{array}{c}
G^b_{eff,+} + N_+ -1\\
N_+
\end{array}\right) 
\left(\begin{array}{c}
G^b_{eff,-} + N_- -1\\
N_-
\end{array}\right). 
\label{counting}
\end{eqnarray}

Consider a grand canonical ensemble
at temperature $T$ and with chemical potential
$\mu_{i} (i=+,-)$ for species $i$, whose partition
function is given by (with $k$ the
Boltzmann constant)

\begin{equation}
Z= \sum_{\{N_{i}\}} W(\{N_{i}\})~
\exp \{\sum_{i} N_{i} (\mu_{i} -\varepsilon_{i})/kT \}~. \label{Z}
\end{equation}
where $W$ is the statistical weight (\ref{counting}). 
One obtains the equations determining the  most 
probable distribution of $n_{\pm}$: 

\begin{equation}
 \sum_{j} \alpha_{ij} n_j(T) = 1 -w_i(T)~,     
\label{mostprobn} 
\end{equation}
with $w_{\pm}(T)$ being determined by the functional
equations
\begin{eqnarray}
{w_{+}}^{\alpha_{++}} (1+w_{+})^{1-\alpha_{++}}
\Bigl( \frac {w_{-}}{1+w_{-}} \Bigr)^{\alpha_{-+}}
& = &  e^{(\varepsilon_{+}-\mu_{+})/kT }~, \nonumber \\
{w_{-}}^{\alpha_{--}} (1+w_{-})^{1-\alpha_{--}}
\Bigl( \frac{w_{+}}{1+w_{+}} \Bigr)^{\alpha_{+-}}
& = &  e^{(\varepsilon_{-}-\mu_{-})/kT }~.             \label{wpm}
\end{eqnarray}
By solving (\ref{mostprobn}), $n_{\pm}$ are expressed in
terms of $w_{\pm}$ as  
\begin{eqnarray}
n_{+}(T) & = & \frac {w_{-}+\alpha_{--}-\alpha_{+-}} {(w_{+}+\alpha_{++})
(w_{-}+\alpha_{--}) -\alpha_{+-}\alpha_{-+} }~, \nonumber \\
n_{-}(T) & = & \frac{w_{+}+\alpha_{++}-\alpha_{-+}} {(w_{+}+\alpha_{++})
(w_{-}+\alpha_{--}) -\alpha_{+-}\alpha_{-+}}~.                \label{npm} 
\end{eqnarray}  

These equations can be solved explicitly as
\begin{eqnarray}
w_+(T) 
&=&\frac{1}{2 f(\nu,m-1)}
\Biggl( -(e^{\Delta/kT}+1) f(\nu,m) \Biggr.\nonumber \\
&&+{\rm sgn}(m) \Biggl.
\sqrt{(e^{\Delta/kT}+1)^2 f(\nu,m)^2-
4 e^{\Delta/kT} f(\nu,m+1) f(\nu,m-1)}\Biggr),\label{wexact1}\\
w_-(T) 
&=&\frac{1}{2 f(\nu,m+1)}
 \Biggl( -(e^{\Delta/kT}+1) f(\nu,m) \Biggr.\nonumber \\
&&-{\rm sgn}(m) \Biggl.
\sqrt{(e^{\Delta/kT}+1)^2 f(\nu,m)^2-
4 e^{\Delta/kT} f(\nu,m+1) f(\nu,m-1)}\Biggr), \label{wexact2}
\end{eqnarray}
where
$f(\nu,m)\equiv  (2pm+1)\nu-m$,  
$
\Delta=\varepsilon_{+}+\varepsilon_-
$
and we have used the charge conservation $\mu_{+}+\mu_{-}=0$.
The sign functions in front of the square roots are 
suitably chosen so that we have 
positive $w_i(T)$ for $\nu \sim \nu_0$.
It is remarkable that a closed expression for $w_i(T)$
is available for our model of FQH systems. This is 
closely related to the structure of the statistics matrix
given by eq. (\ref{jainstat}). 
In Fig.1, $N_\pm(T)$ are plotted against $\nu$.
We have quite remarkably simple form:
$n_{\pm}(T)\bigl|_{\nu=\nu_0}=1/(1+e^{\Delta/2kT})$,
thus we have
\begin{equation}
\rho_{\pm}(T)\Bigl|_{\nu=\nu_0}=
{|e^*_\pm | N_\phi \over V }{1 \over 1+e^{\Delta/2kT}}. \label{rho}
\end{equation}

It may be possible to check experimentally whether our results 
and the fractional charges are correct or not.

Now we proceed to consider thermodynamic
consequences of the above analytic distribution
function. First, the entropy is given by
\begin{eqnarray}
S &=&{ N_{\phi} \over k | 2pm+1|} \sum_{i=\pm}
\left( (1 + n_{i} -
\sum_{j=\pm} \alpha_{ij} {n_{j}})\;
\log (1+ n_i -\sum_{j=\pm} \alpha_{ij} n_{j}) \right.
\nonumber\\
&& \left.- n_i \log n_{i}
- (1- \sum_{j=\pm} \alpha_{ij} n_{j})\;
\log (1 -\sum_{j=\pm} \alpha_{ij} n_{j})\right).
\end{eqnarray}
{}From this expression, we note a duality 
of the system which becomes exact in the 
zero temperature limit. From now on, we 
assume $m>0$ and fix $p$ for simplicity .

Let us study the entropy for the system
with filling in the region 
$m/(2pm+1) \leq \nu \leq (m+1)/[2p(m+1)+1]$
in two ways. 
It can be shown that at low temperatures,
the approximate equality
\begin{equation}
S(m,p) \sim S(m+1,p)
\end{equation}
holds. 
This indicates that at low temperatures,
the same system with $m/(2pm+1) \leq \nu \leq
(m+1)/[2p(m+1)+1]$ can be equivalently  described 
either in terms of quasielectrons in the FQH 
fluid with $\nu(m,p)$ or in terms of quasiholes in the 
FQH fluid with $\nu(m+1,p)$ See Fig. 
Note that at zero temperature we have an 
exact duality $S(m,p)=S(m+1,p)$.

We also note that in between two adjacent
magic fillings $\nu(m,p)$ and $\nu(m+1,p)$, 
there is a filling at which the entropy has a 
local maximum. This filling can be obtained 
by solving 
$
\partial_\nu S(m,p) \bigl|_{T\rightarrow 0}=0.
$
In Table 1, we list some of the fillings 
where the entropy has local maximum. 
Physically, it is believed that the transition between 
plateaus of different quantized Hall conductance 
$\sigma_{xy}$ is due to quantum percolation transition
\cite{DHL} of quasielectrons (or quasiholes from the 
dual point of view). It is natural to expect that the 
entropy should be a maximum at the percolation 
transition point, so one may interpret the solution
of 
$
\partial_\nu S\bigl|_{T\rightarrow 0}=0.
$
as the approximate position
of the transition point, if the percolation transition 
persists in the low disorder limit (see Table 1). 

Second, the thermodynamic potential
$\Omega=-kT\log Z$ is \cite{Wu}

\begin{eqnarray}
\Omega \equiv - PV &=& -kT \sum_{i} G_i\log (1+\frac{n_i}{  1
-\sum_{j} \alpha_{ij}n_j})~   \nonumber\\
&=&
-\frac{kT N_{\phi} }{|2pm+1|}
\log \frac{1+w_+}{w_+}\frac{1+w_-}{w_-}
            \label{Omega}  
\end{eqnarray}
Using the analytic expressions (\ref{wexact1}) 
and (\ref{wexact2}), we
have the following equation of state:
%an expression for the pressure
\begin{eqnarray}
PV &=&\frac{ kT N_{\phi} }{|2pm+1|}
\log \Biggl\{ \frac{(2p\nu-1)}
{e^{\Delta/kT}f(\nu,m+1) f(\nu,m-1) }
\Biggl( \Biggl.
-(1+e^{\Delta/kT}) (2p \nu-1) \Biggr.\label{exactp}\\
&&\Biggl. \Biggl.\quad 
+{\rm sgn}(m)
\sqrt{ (1+e^{\Delta/kT})^2 f(\nu,m)^2 -
4 e^{\Delta/kT} f(\nu,m+1) f(\nu,m-1)    }
\Biggr)   \Biggr\}  .\nonumber
\end{eqnarray}
Examining this expression for the pressure,
we can see how new incompressible 
FQH fluid states arise in thermodynamics:
They emerge at fillings where the pressure
becomes divergent (see Fig. 3). It can be shown that 
$P>0$ in the region $\nu_0(m-1,p)\leq \nu \leq \nu_0(m+1,p)$
and at finite temperatures $P$ diverges at 
temperature-independent fillings
$
\nu =\nu_0(m\pm1,p).
$
The divergence of the pressure, of course, is due to 
the complete filling (of composite fermions) of a 
new Landau level, or the quasiparticle's condensation 
into a new incompressible FQH state, at 
$\nu_0^{\rm new} =(m\pm 1)/(2 p (m\pm 1)+1)$.

Note that $P(m,p)\neq P(m+1,p)$ even at low temperature.
Since the pressures $P(m,p)$ and $P(m+1,p)$ are measured from
different incompressible ground states,
this is not contradictory to the duality of the system.

To conclude, we make the following remarks.
If the filling factor approaches the value 
with the even denominator ($m\rightarrow \infty$)
$
\nu=1/2 p,
$
the statistics matrix degenerates to that of 
free fermion. So, at this filling the 
quasiparticles are nothing but 
free fermions. However, the thermodynamics
for the $\nu=1/2p$ state may not be that simple.
Since if we allow fluctuations of the 
filling fraction near $1/2p$, as in 
a grand canonical ensemble, we have to do 
averaging over many FQH states around the 
filling $1/2p$. It seems important to 
study this averaging effect.

Also, we note that our results crucially depend 
on the assumption (\ref{compfermi}). Though this
seems very natural in the (unprojected)
composite fermion picture, there is numerical 
data for small systems on a sphere \cite{SWYang,ICJ}, 
whose interpretation requires modification of eq. 
(\ref{compfermi}) to
\begin{equation}
G_{eff,\pm}^{(b)}=|N_{\phi,eff}| 
- (N_{\pm}-1) \pm N_\mp.                
\label{compfermi2}
\end{equation}
This will change eq. (\ref{jainstat}) to
$
\alpha_{++}  = 2- \alpha_{--} =
 \alpha_{-+} = - \alpha_{+-}  = 
1+ {2p/ {(2pm+1)}}. $
Despite the numerical evidence in favor of this 
modified parameters, 
it may be possible , for example using (\ref{rho}),
to choose between these from experimental data.

\begin {acknowledgements}
It is a pleasure to thank S. Katsumoto for useful discussions.
The work of Y.S.W. was supported in part by U.S. NSF grant 
PHY-9601277.% and a grant from the Ministry of ****
\end {acknowledgements}

%\end{multicols}{2}

%\end{multicols}

\begin{table}
\begin{tabular}{| r ||  r | r | r | r |}
models with $m \geq 1$, $p=1$ & 
$\nu=0.3692$ & $0.4150$ & $0.4368$ & $0.4497$ \\ \hline
$m \leq -1$, $p=1$& 
 $0.8057$ & $0.6307$ & $0.5850$ &$0.5632$ \\ \hline
$m\geq 1$, $p=2$& 
 $0.2121$ & $0.2267$ & $0.2331$ & $0.2367$\\ \hline
$m\leq -1 $, $p=2$& 
 $0.3064$ & $0.2788$ & $0.2696$  & $0.2648$ \\ 
\end{tabular}
\caption{Some of the transition points obtained from the condition 
$\partial_\nu S\bigl|_{T\rightarrow 0}=0$ are listed. }
\end{table}

\end{document}